\documentclass[12pt]{article}
 \usepackage[dvips]{graphicx}
\usepackage{amssymb}
 \usepackage{amsmath}
 \usepackage{amsxtra}
\begin{document}
\title {On Energy Expenditure per Unit of the Amount of Information }
\author{ A.Granik\thanks{Department of Physics, University of the
Pacific,Stockton,CA.95211; E-mail: agranik@pacific.edu} }
 \maketitle
 \begin{abstract}
It is shown that for an equilibrium state of time-symmetric system
of  non-relativistic strings the energy per unit of information
transfer (storage, processing) obeys the Bekenstein conjecture.
The result is based on a theorem due to A.Kholevo relating the
physical entropy and the amount of information. Interestingly, the
energy in question is the difference between the ensemble averaged
energy  and the Helmholtz free energy.
\end{abstract}

The problem about the energy requirements for the storage,
transfer , and processing of information is one of the most
important problems in the physics of information. In view of the
recent keen interest ( and the attendant very large body of
research work) in possible realizations of quantum computers, the
above problem has direct relevance to quantum systems. Therefore
it seems appropriate to revisit this problem.\\

At the beginning of 1990's B.Schumacher proposed (and later
elaborated) \cite{BS} a conjecture (a generalization of earlier
proposals by Bekenstein \cite{JB} and Pendry \cite{JP}) about an
existence of a quantum limit for the power requirements of a
communication channel. The conjecture relates an amount of
information $H$ conveyed by a quantum channel in a time interval
$\delta t$ and the energy  $E$ required  for the physical
representation of the information in the quantum system. In its
simplest form (proposed for the first time by Bekenstein
\cite{JB}) the conjecture is as follows
\begin{equation}
\label{1} \frac{E}{H}\ge\frac{\hbar}{\delta t}
\end{equation}\\

In what follows we will prove this conjecture ( in a form given by
Eq.\ref{1}) for an equilibrium state of an isolated system. It
will be shown that the energy $E$ turns out to be the difference
between the ensemble averaged energy and the Helmholtz free
energy.\\

To proceed with our reasoning, we consider the partial action
$S_i$ for a time-symmetric system of non-relativistic strings
\cite{SC} of length $\delta$, mass density $\rho$, charge density
$\rho_e$ and tension $\kappa$:
\begin{equation}
\label{2} S_i=\int_0^{\tau_0}d\tau\int_0^{L_0}
\{\frac{1}{2}\rho(\frac{\partial\vec{x_i}}{\partial\tau})^2
-\frac{1}{2}\kappa(\frac{\partial\vec{x_i}}{\partial\lambda})^2-
\rho_e\phi[\vec{x_i}(\tau,\lambda)]\}d\lambda
\end{equation}
Here $\tau$ and ${\lambda}$ are time and space coordinates defined
on the string.\\

If the string is assumed to be  "frozen-in", that is its space
coordinate in the lab frame $\vec{x_i}$ is independent of the
time  $\tau$, that is $\vec{x_i}=\vec{x_i}(\lambda)$, then
 Eq.(\ref{2}) yields:
 \begin{equation}
 \label{3}
 S_i=-\tau_0\int_0^{L_0}\{\frac{1}{2}\kappa(\frac{\partial\vec{x_i}}{\partial\lambda})^2-
\rho_e\phi[\vec{x_i}(\tau,\lambda)]\}d\lambda
\end{equation}
Now we transform this equation into the integral over the energy
of a non-relativistic particle of a mass $m$. This transformation
is possible because, generally speaking, a time interval is not an
absolute concept but is rather defined by an appropriate physical
(periodic) process.\\

Therefore by considering a system of frozen-in strings, we can
safely introduce the respective time interval, say $t$, as follows
\begin{equation}
\label{4} t=\hbar\beta
\end{equation}
where $\beta=1/kT$, $k$ is the Boltzmann constant, and $T$ is the
characteristic  temperature whose lower limit can be taken, for
example, as the Hawking temperature(see \cite{EH})
$$T_H=\frac{1}{\sqrt{\alpha'}}$$
where $\alpha'$ is the string length.\\

This allows us to relate in a very simple fashion the space
variable $\lambda$ to the new time variable $t$ given by (\ref{4})
\begin{equation}
\label{5} t=\frac{\lambda}{c}
\end{equation}
Here $c$ is the speed of light. As a result, Eq.(\ref{3}) becomes
\begin{equation}
\label{6}
S_i=-\tau_0\int_0^{\hbar\beta}cdt\{\frac{1}{2}\frac{\kappa}{c^2}(\frac{d\vec{x_i}}{dt})^2+
\rho_e\phi[\vec{x_i}(t)]\}=-\int_0^{\hbar\beta}dt[\frac{m}{2}(\frac{d\vec{x_i}}{dt})^2+V(t)]
\end{equation}
where $m=\kappa\tau_0/c$ is the mass of a non-relativistic
particle in an external potential $V(t)=\tau_0\rho_e\phi(t)$, as
was indicated Feynman \cite{RF} and later by Chiu \cite{SC}. It
must be noted that in our representation we differ significantly
from both of them, since now the time $t$ is the $\bf{physical}$
time, and not the meta "time" used by Feynman \cite{RF}. The same
comment is true for the expression in figure brackets representing
{\it physical} energy  $E_i$ of a non-relativistic particle of a
mass $m$ at some moment of time $0\le t\le\hbar\beta$
$$E_i=\frac{m}{2}(\frac{d\vec{x_i}}{dt})^2+V(t),$$\\
and not what Feynman  referred as "energy" to remind us
that it was not a physical energy.\\

Now let us consider a system of particles at a thermal equilibrium
at some characteristic temperature (in our case  its lowest limit
can be taken, for example, as the Hawking temperature) and
introduce $\epsilon_i$, the energy averaged over the time interval
$t$ given by (\ref{4}) and(\ref{5})
\begin{equation}
\label{7}
\epsilon_i\equiv\frac{1}{\hbar\beta}\int_0^{\hbar\beta}E_idt
\end{equation} The probability $p_i$ that the system should be in
the same states of energy $\epsilon_i$ as in (\ref{7}) is
\begin{equation}
 \label{8}
 p_i=\frac{e^{-\beta
\epsilon_i}}{\sum_ie^{-\beta \epsilon_i}}\equiv\frac{e^{-\beta
\epsilon_i}}{Z}
\end{equation}

where $$Z=\sum_ie^{-\beta \epsilon_i}$$ is the partition function.\\

On the other hand, the  entropy $\mathcal{S}$ (a $physical$
quantity with a thermodynamic meaning defined for the above
statistical ensemble) is
\begin{equation}
\label{9} \mathcal{S}=-\sum_ip_iLn(p_i)
\end{equation}
Upon substitution of (\ref{8}) in (\ref{9}) we get
\begin{equation}
\label{10}
 \mathcal{S}=\beta <\epsilon>+LnZ
\end{equation}
where the ensemble average $<\epsilon>$ is
$$<\epsilon>=\frac{\sum_i\epsilon_ie^{-\beta\epsilon_i}}{Z}$$

 By using the Kholevo theorem \cite{AK} in its simplest
form $$\mathcal{S}\ge H$$we obtain from (\ref{10})
\begin{equation} \label{10a}
\beta <\epsilon>+LnZ\ge H
\end{equation}
where $H$ is the amount of information. On the other hand, by
definition
\begin{equation}
\label{11}
 LnZ\equiv-F\beta
\end{equation}
where $F$ is the Helmholtz free energy. Therefore (\ref{10a}) yields
\begin{equation}
\label{12} <\epsilon>-F\geqslant\frac{H}{\beta}
\end{equation}
\\

Now using equation (\ref{4}) we express $\beta$ in terms of the
time interval $\delta t$
$$ \beta=\frac{1}{kT}=\frac{\delta t}{\hbar}$$
Inserting this expression in (\ref{12}) we obtain
\begin{equation}
\label{16} \frac{<\epsilon>-F}{H}\ge \frac{\hbar}{\delta t}
\end{equation}\\

If we associate the time interval $\delta t$ with a period of  an
oscillatory process of frequency $\omega$, then inequality
(\ref{16}) becomes:
\begin{equation}
\label{17} \frac{<\epsilon>-F}{H}\ge \hbar\omega
\end{equation}
Since $F$ at constant temperature plays the part of the potential
energy, the difference $<\epsilon>-F$ plays the part of the
average kinetic energy. This means that the minimum average
"kinetic energy" expenditure necessary to transmit, store or
process a unit (a bit, or rather a qubit) of
information is exactly one quantum $\hbar\omega$. \\

Another consequence of inequality (\ref{16}) is that with a
decrease of the temperature, the characteristic time interval
necessary to transmit (store, process), and the respective
temporal rate ( the Kolmogorov information) of information
transmission through the quantum channel decreases. This is easily
explained, since at lower temperatures a system tends to reside at
lower energy states, and its higher states are inaccessible,
unless there is a supply of an additional energy,
which results in an increase of the temperature.\\

\end{document}